\documentclass[aps, 10pt, prd,
               notitlepage, twocolumn, superscriptaddress,
               longbibliography,
               nofootinbib, floatfix]{revtex4-2}

\usepackage[utf8]{inputenc}
\usepackage[T1]{fontenc}
\usepackage{amsmath}
\usepackage{amssymb}
\usepackage{amsfonts}
\usepackage{mathrsfs}
\usepackage{anyfontsize}
\usepackage{verbatim}
\usepackage{CJKutf8}

\usepackage[linktocpage,breaklinks]{hyperref}
\usepackage[dvipsnames]{xcolor}

\usepackage{tensor}
\usepackage{bm}
\usepackage{eucal}
\usepackage[normalem]{ulem}

\usepackage{graphicx}
\usepackage{epsfig}
\usepackage{epstopdf}
\usepackage{natbib}

\hypersetup{colorlinks=true,
            citecolor=NavyBlue,
            linkcolor=NavyBlue,
            urlcolor=NavyBlue}

\usepackage{colortbl}
\usepackage{tabularx}

\newcommand{\dd}{{\rm d}}

\newcommand{\cl}{\textcolor{Black}{({\rm c})}}
\newcommand{\ii}{{\rm i}}
\newcommand{\ee}{{\rm e}}
\newcommand{\lm}{\ell m}

\newcommand{\pd}{\partial}

\newcommand{\calF}{{\cal F}}
\newcommand{\inst}{\textrm{inst}}
\newcommand{\vecm}{\bm{m}}
\newcommand{\sw}{{}_{s}}

\newcommand{\aei}{\affiliation{Max Planck Institute for Gravitational Physics (Albert Einstein Institute),
D-14476 Potsdam, Germany}}
\newcommand{\uiuc}{\affiliation{Department of Physics and Illinois Center for Advanced Studies of the Universe,\\University of Illinois Urbana-Champaign, Urbana, Illinois 61801, USA}}
\newcommand{\umd}{\affiliation{Department of Physics, University of Maryland, College Park, Maryland 20742, USA}}

\begin{document}
\title{Kerr-Newman quasinormal modes and Seiberg-Witten theory}

\begin{abstract}
It was recently suggested the quasinormal-mode spectrum of black holes is
related to a class of four-dimensional $\mathcal{N}=2$ super Yang-Mills theories
described by Seiberg-Witten curves,
a proposal that has been tested for a number of black hole spacetimes.
The aim of this study is to clarify the key ideas of this conjecture to a
non-high-energy-physics audience and test it in a setting that has not yet been
explored: the electromagnetic and gravitational perturbations of Kerr-Newman
black holes in the Dudley-Finley approximation.
In the parameter space we explore, we find numerical evidence that the
conjecture is valid for subextremal black holes and its slowest damped
quasinormal frequencies, thereby providing further support for the conjecture's validity.
In addition, we exploit the symmetries of the four-dimensional $\mathcal{N}=2$
super Yang-Mills theory to obtain a strikingly simple isospectral version of the radial
Dudley-Finley equation.
\end{abstract}

\begin{CJK*}{UTF8}{mj}
\author{Hector O. Silva}          \aei \uiuc
\author{Jung-Wook Kim (김정욱)}   \aei
\author{M. V. S. Saketh}          \aei \umd
\maketitle
\end{CJK*}

\section{Introduction}
\label{sec:intro}

The response of black holes to linear perturbations exhibits a characteristic
``ringdown,'' that is dominated by a linear superposition of complex-valued
frequencies known as quasinormal mode frequencies~\cite{Vishveshwara:1970zz,Press:1971wr}.
The computation of black hole quasinormal modes is usually formulated as a
spectral problem: solving for the eigenvalues of a (set of) differential
equation(s) with suitable boundary conditions~\cite{Chandrasekhar:1975zza}.

Recently, a proposal by Aminov et~al.~\cite{Aminov:2020yma} has stirred a lot of interest.
In this proposal, the gravitational problem of studying quasinormal modes is addressed from a gauge-theoretic approach to spectral problems,
adding another example to the list of intriguing relations between gauge theory and gravity.
This approach to quasinormal modes has been tested in various spacetime geometries~\cite{Bianchi:2021xpr,Bianchi:2021mft,Fioravanti:2021dce,Fioravanti:2022bqf,Bianchi:2023sfs,Bianchi:2023rlt}
and has also been applied to superradiance~\cite{Ge:2024jdx,Cipriani:2024ygw}.
Moreover,
the approach was extended to the construction of eigenfunctions from two-dimensional conformal field theories,
based on the Alday-Gaiotto-Tachikawa correspondence~\cite{Alday:2009aq} in Ref.~\cite{Bonelli:2021uvf};
see also Refs.~\cite{Bianchi:2021mft,Consoli:2022eey,Aminov:2023jve,BarraganAmado:2023apy,Lei:2023mqx,Fucito:2023afe,Bautista:2023sdf,Fucito:2024wlg,Bianchi:2024mlq,Aminov:2024mul,Bautista:2024emt,Matone:2024ytm,Cipriani:2025ikx,Arnaudo:2025kof} for further applications.
Our aim here is to clarify the key ideas of the proposal put forward in Ref.~\cite{Aminov:2020yma}
to an audience not familiar with Seiberg-Witten theory~\cite{Seiberg:1994rs,Seiberg:1994aj}
and test it in a setup that has not yet been explored: the gravitoelectromagnetic perturbations of the Kerr-Newman spacetime.

The coupled gravitoelectromagnetic perturbations of the Kerr-Newman solution
are not separable when decomposed in modes, but they are reducible to a
coupled system of partial differential equations, as shown by Chandrasekhar~\cite{Chandrasekhar:1978RSPSA.358..421C}.
This ``apparent indissolubility of the coupling between spin-1 and spin-2 fields''~\cite{Chandrasekhar:1985kt},
has prevented an analysis of the quasinormal-mode spectrum of the Kerr-Newman solution
until the works by Dias et al.~\cite{Dias:2015wqa,Dias:2021yju,Dias:2022oqm},
who tackled the mode calculations directly from the system of coupled partial differential equations.
However, separability is possible in some limits, for example, by working perturbatively
in small values of black hole spin~\cite{Pani:2013ija,Pani:2013wsa,Blazquez-Salcedo:2022eik}
or charge-to-mass ratio~\cite{Mark:2014aja}.
Here, we will use an approximation introduced by Dudley and Finley~\cite{Dudley:1977zz,Dudley:1978vd}
that results in a deformation to the Teukolsky equation~\cite{Teukolsky:1973ha}. The resulting Dudley-Finley equation is amenable
to analysis using the techniques offered by the gauge-theoretic approach to quasinormal-mode computations~\cite{Aminov:2020yma}.

This paper is organized as follows. In Sec.~\ref{sec:sw_ash_conj}, we present a
summary of Seiberg-Witten theory and the Aminov-Grassi-Hatsuda
conjecture. In Sec.~\ref{sec:per_kn}, we review the Dudley-Finley equation, express
it in its canonical form, derive a dictionary with four-dimensional $\mathcal{N}=2$~SU(2)
super Yang-Mills theory, and review how quasinormal frequencies are determined
using continued fractions, a standard approach in the gravitational physics literature.
Then, in Sec.~\ref{sec:nr}, we compare our numerical calculations of the
quasinormal frequencies using gauge-theoretical and continued-fraction methods.
In Sec.~\ref{sec:hid_sym}, we explore the hidden symmetries of the Dudley-Finley
equation, and use them to obtain a remarkably simple, and isospectral, version
thereof.
We summarize our findings in Sec.~\ref{sec:conclusions}, and indicate
directions for future work.

For the gravitational equations, we use the mostly plus metric signature,
units where $c = G = 1$, and set $2M=1$, where $M$ is the black hole's mass.
For the Seiberg-Witten equations, we use $\hbar = 1$.

\section{Seiberg-Witten theory and the Aminov-Grassi-Hatsuda conjecture}
\label{sec:sw_ash_conj}

The Seiberg-Witten theory~\cite{Seiberg:1994rs,Seiberg:1994aj} is an attempt to explain confinement in quantum chromodynamics (QCD) using supersymmetry.
Being an asymptotically free theory, QCD is described by different degrees of freedom at high energy scales (UV) and low energy scales (IR).
In the UV, QCD is weakly coupled and described by quarks and gluons, which are the fundamental degrees of freedom of the theory.
In the IR, QCD is strongly coupled and described by mesons and baryons, which are effective degrees of freedom of the theory.
In Seiberg-Witten theory, confinement is understood as condensation of monopoles where the (non-Abelian) electric fields become flux tubes due to the dual Meissner effect.
The theory provides tools to compute the masses of the particles in the IR description of the theory from its UV description and, in particular, the theory has monopoles in the IR.
Confinement is argued by showing that the theory admits a vacuum solution where the monopoles become massless, leading to the magnetic dual of superconductivity. See Ref.~\cite{Tachikawa:2013kta} for an introduction to the subject.

The Seiberg-Witten curve describes the IR particle spectrum of a (four-dimensional) $\mathcal{N} = 2$ gauge theory in the language of algebraic geometry.
The Seiberg-Witten curve is a complex one-dimensional manifold, which is given as the set of solutions to a polynomial of complex variables $x$ and $p$,
\begin{align} \label{eq:SWC_classical}
F (p,x ; E) &= 0 \,,
\end{align}
where $E$ parametrizes the possible vacuum solutions (the vacuum moduli space) of the theory.\footnote{To simplify the Introduction we ignored the dependence on $\Lambda_{N_f}$, the analogue of $\Lambda_{\text{QCD}}$ from QCD, which characterizes the energy scale where the theory becomes strongly coupled. We also ignored contributions from matter fields, for instance, the fundamental hypermultiplets considered in Ref.~\cite{Aminov:2020yma}, which are the analogues of quarks in QCD.}
For simplicity, we assume the curve to have genus $1$ (i.e., the topology of the torus ${T}^2 = S^1 \times S^1$), which is the case for the gauge group SU(2).
The curve has two cycles (noncontractible closed paths), which we label as $A$ and $B$.

The (classical) periods $\Pi_{A,B}^{\cl}$ are defined as integrals over the corresponding cycles,\footnote{Some definitions were modified from Ref.~\cite{Aminov:2020yma} for coherence.}
\begin{align}
\begin{aligned}
\Pi_A^{\cl} &= 2 \pi \ii \mathfrak{a}^{\cl} = \oint_A p \, \dd x \,,
\\ \Pi_B^{\cl} &= \mathfrak{a}_D^{\cl} = \frac{\partial \mathcal{F}}{\partial \mathfrak{a}^{\cl}} = \oint_B p \, \dd x \,,
\end{aligned} \label{eq:period_cl}
\end{align}
which compute the (semiclassical) mass of the $W$ bosons ($|\mathfrak{a}^{\cl}|$) and magnetic monopoles ($|\mathfrak{a}_D^{\cl}|$).
They constitute the Bogomol'nyi-Prasad-Sommerfield (BPS) particle spectrum of the theory in the IR.
The differential form $p \, \dd x$ is known as the Seiberg-Witten differential, and the function $\mathcal{F}$ appearing in $\mathfrak{a}_D^{\cl}$ is called the prepotential.

In general, the masses acquire quantum corrections, and we denote the quantum version of the variables by omitting the superscript $\cl$.
These quantum corrections are understood as instanton effects in the UV description of the theory.
In this approach, $\mathfrak{a}$, known as the Coulomb vacuum expectation value (VEV), is an independent variable, and quantum corrections to $\mathfrak{a}_D = \partial_{\mathfrak{a}} \calF$ are computed.
The prepotential $\calF$ is identified with the Nekrasov-Shatashvili free energy $\calF^{\,(N_f)}$~\cite{Nekrasov:2009rc}, which is computed from a combinatoric formula obtained through supersymmetric localization techniques~\cite{Nekrasov:2002qd}.
The resulting expression for $\mathfrak{a}_D$ is a perturbative series in $\Lambda_{N_f}$, where the power of $\Lambda_{N_f}$ is related to the number of instantons used in supersymmetric localization.

In the IR, the quantum corrections are computed as corrections to the periods $\Pi_{A,B}$ due to quantization of the Seiberg-Witten curve,\footnote{Quantization may involve variable redefinitions of the classical Seiberg-Witten curve~\eqref{eq:SWC_classical}. The notation $\tilde{F}$ reflects this possibility.}
\begin{align}
\tilde{F} (\hat{p} , x; E) &= 0 \,,\quad \hat{p} = - \ii \, \partial_x \,,
\label{eq:sw_curve}
\end{align}
which is understood as the quantization of a classical integrable system corresponding to the curve.
This viewpoint is known as the Bethe/Gauge correspondence~\cite{Nekrasov:2009rc}; see also, Refs.~\cite{Gorsky:1995zq,Mironov:2009uv}.
As an operator equation, Eq.~\eqref{eq:sw_curve} only makes sense when acting on a test wave function $\psi (x)$,
\begin{align} \label{eq:SWC_quantum}
\tilde{F} (\hat{p} , x; E) \, \psi (x) &=0 \,,\quad \psi (x) = \exp \left(\ii \int^x P \, \dd x' \right) \,,
\end{align}
where $P(x; E)$ is the quantum-corrected $p(x; E)$ variable.
The periods~\eqref{eq:period_cl} become
\begin{align}
\Pi_A &= 2 \pi \ii \mathfrak{a} = \oint_A P \, \dd x \,,\quad \Pi_B = \mathfrak{a}_D = \oint_B P \, \dd x \,,
\end{align}
after quantization; recall we have set $\hbar = 1$.
The periods can be viewed as complexified action variables in this context.

The key idea of Ref.~\cite{Aminov:2020yma} is based on two observations:
(i) the periods $\Pi_{A,B}$ can be computed in the UV description of the theory as instanton series based on supersymmetric localization techniques, and
(ii) the quantum Seiberg-Witten curve~\eqref{eq:SWC_quantum} solves an eigenvalue problem of 
$E$ when the Bohr-Sommerfeld quantization condition is imposed on the quantum periods $\Pi_{A,B}$,
\begin{align} \label{eq:BS_quant_cond}
\Pi_I &= 2 \pi
\left( n + \frac{1}{2} \right) \,,\quad n \in \mathbb{Z} \,,
\end{align}
where the choice of the period $I = A, B$ depends on the domain of $x$ in which we wish to solve the differential equation~\eqref{eq:SWC_quantum} on.
Therefore, an eigenvalue problem of a differential equation having the form Eq.~\eqref{eq:SWC_quantum} can be converted to an algebraic root-finding problem [Eq.~\eqref{eq:BS_quant_cond}] provided that we know the UV description of the gauge theory leading to the quantum Seiberg-Witten curve Eq.~\eqref{eq:SWC_quantum}.

The main claim of Ref.~\cite{Aminov:2020yma} is that the eigenvalue problem for a differential equation that can be reduced to the form
\begin{align} \label{eq:AGH_SWC}
\left[ \frac{\dd^2}{\dd z^2} +
\frac{1}{z^2 (z - 1)^2} \sum_{i=0}^4 \hat{A}_i z^i \right] \Psi (z) &= 0 \,,
\end{align}
can be solved as an algebraic root-finding problem using four-dimensional $\mathcal{N} = 2$ SU(2) super Yang-Mills theory with $N_f = 3$ fundamental hypermultiplets, since this gauge theory has a quantum Seiberg-Witten curve that can be converted to the form Eq.~\eqref{eq:AGH_SWC}.
The coefficients $\hat{A}_i$ as a function of IR variables $\{ E , \vecm , \Lambda_3 \}$ are given by [see~Ref.~\cite{Aminov:2020yma},~Eq.~(2.29)]
\begin{align}\label{eq:Atilde_coefficients}
\begin{aligned}
    \hat{A}_0 &= - \tfrac{1}{4}(m_1 - m_2)^2 + \tfrac{1}{4} \,, \\
    \hat{A}_1 &= - E - m_1 m_2 - \tfrac{1}{8} m_3 \Lambda_3 - \tfrac{1}{4} \,, \\
    \hat{A}_2 &= E + \tfrac{3}{8} m_3 \Lambda_3 - \tfrac{1}{64} \Lambda_3^2 + \tfrac{1}{4} \,, \\
    \hat{A}_3 &= - \tfrac{1}{4} m_3 \Lambda_3 + \tfrac{1}{32} \Lambda_3^2 \,, \\
    \hat{A}_4 &= - \tfrac{1}{64} \Lambda_3^2 \,,
\end{aligned}
\end{align}
where $\vecm =\{ m_1 , m_2 , m_3 \}$ is the mass vector.
One concrete application of this approach is using the quantum Seiberg-Witten curve to compute quasinormal modes of black holes~\cite{Aminov:2020yma}, since the governing equation of linearized perturbations on black hole backgrounds can be converted to the form Eq.~\eqref{eq:AGH_SWC}.

One necessary intermediate step is the translation of the UV variables into IR variables, since the independent variables of the two descriptions are not the same.
The Coulomb VEV $\mathfrak{a}$ is an independent variable in the UV description, but a dependent variable in the IR description.
The independent IR variable corresponding to $\mathfrak{a}$ is $E$, and the relation between the two variables is called the Matone relation~\cite{Matone:1995rx,Flume:2004rp}
\begin{align} \label{eq:Matone1}
E &= \mathfrak{a}^2 - \frac{\Lambda_{N_f}}{4 - N_f} \frac{\partial \mathcal{F}_{\text{inst}}^{(N_f)} (\mathfrak{a} , \vecm , \Lambda_{N_f}) }{\partial \Lambda_{N_f}} \,,
\end{align}
which is presented as Eq.~(A.16) in Ref.~\cite{Aminov:2020yma}.
This relation can be inverted to $\mathfrak{a} = \mathfrak{a} (E, \vecm, \Lambda_{N_f})$ as a perturbative series in $\Lambda_{N_f}$, which corresponds to the period $\Pi_A = 2 \pi i \mathfrak{a}$ computed using IR variables.\footnote{We choose the branch $\mathfrak{a} = + \sqrt{E} + \cdots $ for the inversion, but the other branch choice $\mathfrak{a} = - \sqrt{E} + \cdots $ is also a valid option, which results in an extra minus sign for the period $\Pi_B$.}
For the period $\Pi_B = \mathfrak{a}_D = \partial_\mathfrak{a} \calF^{\,(N_f)}$, we use Eq.~(A.15) of Ref.~\cite{Aminov:2020yma} and refer the reader to the same reference for its computation.\footnote{This computation involves the logarithm of the gamma functions that should be computed using the \emph{log gamma} function to avoid branch-cut ambiguities. This function is implemented in various packages, e.g., as \texttt{LogGamma} in \texttt{Mathematica} or \texttt{loggamma} in \texttt{SciPy}.}

The period to be quantized from Eq.~\eqref{eq:BS_quant_cond} can be determined from the classical limit where $E > 0$ is the most dominant variable.
In this limit, the quantum Seiberg-Witten curve~\eqref{eq:AGH_SWC} is approximated by
\begin{align}
- p(z)^2 + \frac{E}{z(z-1)} \simeq 0 \quad \Rightarrow \quad p (z) \simeq \sqrt{\frac{E}{z(z-1)}} \,,
\end{align}
where we place the branch cut between $z=0 $ and $z=1$.
The period for the cycle encircling the branch points $z = 0$ and $z=1$ is approximated as
\begin{align}
\int_0^1 p(x + \ii 0^+) \, \dd x + \int_1^0 p(x - \ii 0^+) \, \dd x \simeq 2 \pi \ii \sqrt{E} \,,
\end{align}
where $x$ is real. From the Matone relation~\eqref{eq:Matone1}, we find $\Pi_A = 2 \pi \ii \mathfrak{a} \simeq 2 \pi \ii \sqrt{E}$; thus, we quantize $\Pi_A$ when the relevant domain is $z \in [0,1]$.
The period for the cycle encircling the branch point $z=1$ and the asymptotic branch point
$z = R \simeq {16 \sqrt{2E}} / {\Lambda_3} \gg 1$
is approximated as\footnote{This branch cut is due to
$\hat{A}_4 = - \frac{\Lambda_3^2}{64}$, where we assume $\Lambda_3 > 0$.}
\begin{align}
\begin{aligned}
& \int_1^R \sqrt{\frac{E}{x(x-1)}} \, \dd x + \int_R^1 \left[ - \sqrt{\frac{E}{x(x-1)}} \, \right] \dd x
\\ &\quad = 2 \sqrt{E} \log (4R) \simeq \sqrt{E} \log \left( \frac{E}{\Lambda_3^2} \right) \,,
\end{aligned}
\end{align}
where the extra minus sign in the second integral is due to crossing the branch cut and picking up the values on the second Riemann sheet.
This matches the period $\Pi_B = \partial_{\mathfrak{a}} \calF \simeq \sqrt{E} \log \left( {E}/{\Lambda_3^2} \right)$ in the same limit; therefore, we quantize $\Pi_B$ when the relevant domain is $z \in [1,\infty)$.

In a similar vein, Ref.~\cite{Aminov:2020yma} also proposes that the eigenvalue problem for a differential equation of the form,
\begin{align} \label{eq:AGH_SWC_2}
\left[ \frac{\dd^2}{\dd z^2} +
\frac{1}{z^4} \sum_{i=0}^4 \tilde{A}_i z^i \right] \Psi (z) &= 0 \,,
\end{align}
can be solved as an algebraic root-finding problem using four-dimensional $\mathcal{N} = 2$ SU(2) super Yang-Mills theory with $N_f = 2$ fundamental hypermultiplets.
The coefficients $\tilde{A}_i$ can be found in Ref.~\cite{Aminov:2020yma}, Eq.~(2.31).
To determine the period to be quantized, we evaluate the classical period between the branch points $z = \tilde{R}^{-1}, \tilde{R}$, where $
\tilde{R} \simeq {4\sqrt{E}} / {\Lambda_2} \gg 1$.
\begin{align}
\begin{aligned}
2 \int_{\tilde{R}^{-1}}^{\tilde{R}} \frac{\sqrt{E}}{x} \, \dd x
\simeq 2 \sqrt{E} \log \left( \frac{E}{\Lambda_2^2} \right) \,.
\end{aligned}
\end{align}
Similar to the $N_f = 3$ case, this matches the period $\Pi_B = \partial_{\mathfrak{a}} \calF \simeq 2 \sqrt{E} \log \left( {E} / {\Lambda_2^2} \right)$ in the same limit; we quantize $\Pi_B$ when the relevant domain is $z \in [0,\infty)$.

While its mathematical foundation, the Bethe/Gauge correspondence, is firmly established~\cite{Jeong:2018qpc,Grassi:2019coc},
the proposal of Ref.~\cite{Aminov:2020yma} should be considered conjectural or incomplete.
This is because of the difficulty of identifying the correct period to be quantized by the Bohr-Sommerfeld quantization condition~\eqref{eq:BS_quant_cond} without further input.\footnote{It is possible to construct the wave function $\Psi(x)$ of Eq.~\eqref{eq:AGH_SWC} from gauge theory as a two-dimensional conformal field theory correlation function~\cite{Bonelli:2021uvf}, which can be used to relate Bohr-Sommerfeld quantization conditions to the boundary conditions~\cite{Ge:2024jdx}.}

The Seiberg-Witten curve was motivated from monodromy properties of $\mathfrak{a}$ and $\mathfrak{a}_D$ as a function of $E$~\cite{Seiberg:1994rs}; they are multivalued in $E$ and generally do not return to their original values when $E$ moves on the complex plane and returns to itself.
The periods $\Pi_{A,B}$ also have this property, which is inherited from their defining cycles.
A concrete example given in the original Seiberg-Witten paper~\cite{Seiberg:1994rs} is the cycle $B$ becoming the linear combination of cycles $B' = 2 A - B$.
In this case, the period $\Pi_B$ becomes $\Pi_{B'} = 2 \Pi_A  - \Pi_B$, or $\mathfrak{a}_D \to - \mathfrak{a}_D + 4 \pi \ii \mathfrak{a}$.

This implies that although we have identified the period to be quantized from analyzing the classical limit of the curve Eq.~\eqref{eq:AGH_SWC}, it may not be the correct period to be quantized, as the parameters are complex valued and monodromy ambiguity will affect the periods.
For example, it may turn out that the correct period to be quantized is $\Pi_B + \Pi_A$ for some choice of parameters, even though we are interested in the domain $z \in [0 , \infty)$.

A related phenomenon is the dictionary ambiguity, where multiple choices of the Seiberg-Witten parameters yield the same quantum Seiberg-Witten curve.
A concrete example is the sign ambiguity of the first two mass parameters $\pm m_{1,2}$, where both sign choices lead to the same quantum Seiberg-Witten curve, since the $\hat{A}_i$ coefficients~\eqref{eq:Atilde_coefficients} of Eq.~\eqref{eq:AGH_SWC} only depends on the combinations $(m_1 - m_2)^2$ and $m_1 m_2$.
The dependence of quantum Seiberg-Witten curve's boundary conditions on the choice of the dictionary has been studied in Refs.~\cite{Casals:2021ugr,Ge:2024jdx}.

Before concluding the overview on quantum Seiberg-Witten curves and how they can be used to obtain quasinormal modes of black holes, let us comment on the relation between this approach and the Wentzel-Kramers-Brillouin (WKB) approach.
After all, WKB methods have been used to this very same problem since the 1980s~\cite{Schutz:1985km,Iyer:1986np}.
The Bohr-Sommerfeld quantization condition
\eqref{eq:BS_quant_cond}
should be familiar to a reader with knowledge in WKB methods.
In fact, we arrive at the same differential equations when the quantum Seiberg-Witten curve
\eqref{eq:SWC_quantum}
is chosen to have the form $\tilde{F} = - \hat{p}^2 + Q(x) = {\dd^2}/{\dd x^2} + Q(x)$.
This is not a coincidence: one way of demonstrating the Bethe/Gauge correspondence is to compute the quantum corrections \emph{using} (exact) WKB methods applied to the quantum Seiberg-Witten curve and showing that they match the instanton calculations~\cite{Mironov:2009uv,Mironov:2009dv}.
The approach of using quantum Seiberg-Witten curves to compute quasinormal modes can be viewed as reversing the flow of logic; we solve the (exact) WKB problem using instanton calculus, based on other demonstrations of the Bethe/Gauge correspondence.
It has been argued that instanton calculus has an advantage over the WKB approach since the former is exact in $\hbar$ and expected to be a convergent expansion in the instanton counting parameter $\Lambda_{N_f}$ (in the semiclassical regime $|\Lambda_{N_f}/\mathfrak{a}| \ll 1$~\cite{Grassi:2019coc}), while the latter is known to be an asymptotic series in the formal expansion parameter $\hbar$ having vanishing radius of convergence~\cite{Aminov:2020yma}.

We show in subsequent sections that linearized vector and tensor perturbations on Kerr-Newman backgrounds can be converted to the form Eq.~\eqref{eq:AGH_SWC} in the Dudley-Finley approximation; therefore, the problem can be approached using quantum Seiberg-Witten curves.
We present a parameter dictionary that is consistent with the period choice based on the classical limit, study the numerical convergence of the Seiberg-Witten approach to quasinormal modes, and leave the study of period ambiguities/dictionary ambiguities to the future.

\section{Perturbations of Kerr-Newman black holes}
\label{sec:per_kn}

In light of the current status of the correspondence between black hole
quasinormal modes and Seiberg-Witten theory proposed in Ref.~\cite{Aminov:2020yma},
it is reasonable to study whether the correspondence holds in different examples.
In this section, we present the example we will analyze the spin-weighted
perturbations of the Kerr-Newman solution in the Dudley-Finley
approximation~\cite{Dudley:1977zz,Dudley:1978vd}.

\subsection{The Dudley-Finley equation} \label{sec:DFeq}

Dudley and Finley~\cite{Dudley:1977zz,Dudley:1978vd} studied the separability of the
linear perturbations of the family of solutions of the Einstein-Maxwell theory obtained by Pleba\'{n}ski
and Damia\'{n}ski~\cite{Plebanski:1976gy}; see also Ref.~\cite{Griffiths:2005qp} or Sec.~21.1.2 of the monograph~\cite{Stephani:2003tm}.
This family of solutions includes all electrovacuum spacetimes of Petrov-type D,
of which the Kerr-Newman metric~\cite{Newman:1965my} is its most famous specimen.
The coupled gravitoelectromagnetic perturbations of the Kerr-Newman solution
are not separable when decomposed in modes; see Ref.~\cite{Chandrasekhar:1985kt}, Sec.~111,
or Ref.~\cite{Giorgi:2020ujd}, Sec.~1, for a discussion.\footnote{This poses challenges in proving
the linear stability of the Kerr-Newman solution. However, see Ref.~\cite{Giorgi:2020ujd}
for an alternative approach to address this problem based on physical-space methods.}
Nonetheless, Refs.~\cite{Dudley:1977zz,Dudley:1978vd}, showed that mode separability is
possible if either the background metric or the electric field are kept fixed.
The resulting equation that describes the radial dependence of the
perturbations is a deformation of the Teukolsky equation~\cite{Teukolsky:1973ha}
known as the Dudley-Finley equation. The angular dependence of the
perturbations is described by the spin-weighted spheroidal harmonics.

The Dudley-Finley equation is
\begin{align}
\label{eq:DFeq}
\begin{aligned}
&\Delta^{-s} \frac{\dd}{\dd r} \left[ \Delta^{s+1} \frac{\dd \, \sw R_{\lm}}{\dd r} \right]
+ \frac{1}{\Delta} \left[ K^2 - \ii s \frac{\dd \Delta}{\dd r} K \right.
\\
&\quad
\left.
+ \, \Delta \left( 2 \ii s \frac{\dd K}{\dd r} - \lambda_{\lm} \right)
\right] \, \sw R_{\lm} = 0 \,,
\end{aligned}
\end{align}
where we defined
\begin{equation}
    K = (r^2 + a^2) \, \omega - a m\,,
    \quad
    \Delta = (r - r_{+})(r - r_{-}) \,,
\end{equation}
and
\begin{equation}
    r_\pm = (1 \pm b) / 2 \,, \quad \textrm{where} \quad b = \sqrt{1 - 4 (a^2 + Q^2)} \,,
\label{eq:rpm}
\end{equation}
are the locations of the outer ($r_{+}$) and inner ($r_{-}$) horizons in Boyer-Lindquist coordinates, and
$\sw\lambda_{\lm} = \sw A_{\lm} + (a \omega)^2 - 2 am \omega$.
Here, $\sw A_{\lm}$ is a separation constant and it corresponds to the
eigenvalue of the spin-weighted spheroidal harmonic equation;
see~Eq.~\eqref{eq:S_harmonic_eq} next.
The spin-weight parameter $s$ has values $0$, $-1$ and $-2$ for scalar,
electromagnetic, and gravitational perturbations, respectively.
The black hole's angular momentum per unit mass is given by $a$, and is bound
to the interval,
\begin{equation}
    0 \leq a < \sqrt{1 - 4 Q^2} /  2\,,
\end{equation}
in our units, $2 M  = 1$. The Dudley-Finley equation is exact only when $s = 0$ or $Q = 0$.
The former case describes the perturbations of a massless scalar field to the Kerr-Newman
background, whereas in the latter case, we recover the Teukolsky equation~\cite{Teukolsky:1973ha}.

In general, when $a \neq 0$, the separation constant $\sw A_{\lm}$ has to
be determined, for a given value of $\omega$, by solving the equation,
\begin{align} \label{eq:S_harmonic_eq}
    &\frac{\dd}{\dd u} \left[ (1 - u^2) \frac{\dd \, \sw S_{\lm}}{\dd u} \right]
    + \biggl[ (a \omega u)^2 - 2 a \omega s u + s + \sw A_{\lm} \phantom{\frac{}{}}  \nonumber \\
    &\qquad - \frac{(m + su)^2}{1-u^2} \biggr] \, \sw S_{\lm} = 0 \,,
\end{align}
where $u = \cos \vartheta$, related to the polar angle
$\vartheta$ of Boyer-Lindquist coordinates, and imposing
boundary conditions such that the eigenfunctions $\sw S_{\lm}$ are finite at the
regular points $u = \pm 1$.
These eigenfunctions are known as the spin-weighted spheroidal harmonics, and
$c = a\omega$ is the spheroidicity parameter $c$, which is complex
valued in general.
For vanishing spheroidicity, $\sw S_{\lm}$ become the spin-weighted
spherical harmonics, with eigenvalues,
\begin{equation}
    \sw A_{\lm} = \ell ( \ell + 1) - s (s + 1)\,,
    \quad \textrm{for} \quad c = 0.
\end{equation}
Analytical corrections in $c$ to the foregoing equation can be obtained
perturbatively; see, for instance Refs.~\cite{Press:1973zz,Fackerell:1977,Seidel:1988ue,Berti:2005gp}.

Before proceeding, one may ask: what is the regime of validity of the
Dudley-Finley approximation?
This question was studied by Berti and Kokkotas in the nonrotating limit
of the Dudley-Finley equation~\cite{Berti:2005eb}.
In this limit, they compared the quasinormal frequencies obtained from the
Dudley-Finley equation against the respective frequencies obtained from a
perturbed Reissner-Nordstr\"om black hole where one ``freezes'' either the
metric or electromagnetic perturbations.

To provide some context, we recall that it is known from the works of Zerilli~\cite{Zerilli:1974ai}, Moncrief~\cite{Moncrief:1974gw,Moncrief:1974ng,Moncrief:1975sb},
and Chandrasekhar and Xanthopoulos~\cite{Chandarsekhar:1979RSPSA.367....1C,Xanthopoulos:1981RSPSA.378...73X}
that the coupled linear metric and electromagnetic perturbations of the Reissner-Nordstr\"om solution are separable.
Specifically, the perturbations of axial and polar parities can each be reduced
to a pair of coupled differential equations that describe the interaction
between gravitational and electromagnetic perturbations.
Each pair of equations can then be decoupled by introducing a suitable linear combination
of the original perturbations variables.
Remarkably, Chandrasekhar showed that
each decoupled equation describing perturbations of one parity have the same
quasinormal mode spectra as its counterpart describing perturbations of the
other parity~\cite{Chandrasekhar:1980RSPSA.369..425C}.
For this reason, calculations of the quasinormal frequencies of the Reissner-Nordstr\"om
solution often make use of the decoupled equations of axial parity, which are simpler
in form.

In Ref.~\cite{Berti:2005eb}, Berti and Kokkotas took the \emph{coupled}
equations of axial parity and imposed on them that either the electromagnetic
or gravitational perturbation vanished.
They found that the quasinormal frequencies of the resultant equations agree
well with the respective results obtained using the Dudley-Finley equation for
$s=-1$ and $-2$ as long as $Q \lesssim 1/4$.
Surprisingly, the agreement between the two approximations is not exact. This
suggests that the approach of Dudley and Finley is not equivalent to simply
fixing either the background metric or electric field~\cite{Berti:2005eb}.
Nonetheless, Ref.~\cite{Berti:2005eb} also found that the Dudley-Finley equation
reproduces within 1\% the quasinormal frequencies of the coupled system of
equations when $Q \lesssim 1/4$, and argued that the same level of accuracy
would hold for the Kerr-Newman solution as long as the black hole spin is
sufficiently small; see also Ref.~\cite{Pani:2013wsa}.
However, Mark et al.~\cite{Mark:2014aja} found that the Dudley-Finley equation
no longer predicts the quasinormal-frequencies accurately as the spin
increases, with the exception of a special set of modes of rapidly rotating
black holes.

\subsection{Canonical form and Seiberg-Witten dictionary}
\label{sec:canonical_form}

Having presented the Dudley-Finley equation and discussed its regime of
validity, we know establish its connection to Seiberg-Witten theory.
The Dudley-Finley equation~\eqref{eq:DFeq} can be rewritten in the
form Eq.~\eqref{eq:AGH_SWC} through the redefinition
\begin{equation}\label{eq:DF2SWC_1}
    \sw R_{\lm} = \Delta^{-(s+1)/2} \, \sw\Psi_{\lm} \,,
\end{equation}
together with the change of coordinates
\begin{equation} \label{eq:r_to_z}
    r = r_+ + (r_+ - r_-) (z-1) \,.
\end{equation}
The $A_i$ coefficients corresponding to the $\hat{A}_i$ coefficients of Eq.~\eqref{eq:AGH_SWC} are
\begin{equation*}
    A_0 = \frac{1}{4} + \frac{\left[ 2am - \ii(r_+-r_-) s + 2 (Q^2 - r_-) \, \omega \right]^2}{4 (r_+ - r_-)^2} \,,
\end{equation*}
\begingroup
\allowdisplaybreaks
\begin{align}\label{eq:DF2SWC_2}
\begin{aligned}
    A_1 &= \sw A_{\ell m} + s(s+1) - 2 (1 - Q^2) \, \omega^2 \\
        &\quad + \frac{1}{r_+ - r_-} \left\{
        \left[2 - 6 Q^2 - (2r_- + 3) a^2 \right] \omega^2
        \right. \\
        &\quad
        \left.
        - 2 \left[ a m + \ii s (2a^2 + Q^2) \right] \omega + 2 \ii ams
        \right\} \,,
    \\
    A_2 &= - \sw A_{\ell m} - s(s+1) - 3 \ii s (r_+ -r_-) \, \omega \\
    &\quad + (6 r_- -5 a^2 - 6 Q^2 ) \, \omega^2\,,
    \\
    A_3 &= 2 \, (\ii s + 2 r_- \omega) \, (r_+ - r_-) \, \omega \,,
    \\
    A_4 &= (r_+ - r_-)^2 \, \omega^2 \,.
\end{aligned}
\end{align}
\endgroup

We can now relate the Seiberg-Witten parameters $\{ E , \vecm , \Lambda_3 \}$
to the Dudley-Finley-equation parameters $\{a, Q, \omega, m, s, \sw A_{\lm} \}$ by matching the coefficients
$A_{i}$, listed in Eq.~\eqref{eq:DF2SWC_2}, with the coefficients $\hat{A}_{i}$, listed in Eq.~\eqref{eq:Atilde_coefficients}.
One possible solution is
\begin{align}
\begin{aligned}
\Lambda_3 &= - 8 \ii (r_+ - r_-) \omega \,,
\\ E &= - \tfrac{1}{4} - \sw A_{\ell m} - s(s+1) + ( 2 - a^2 - 2 Q^2 )\,\omega^2 \,,
\\ m_1 &= - s - \ii \omega \,,\quad m_3 = s - \ii \omega \,,
\\ m_2 &= {\ii \left[ 2 a m - (1 - 2Q^2)\,\omega \right]} / (r_+ - r_-) \,.
\end{aligned} \label{eq:DF2SWC_dic1}
\end{align}

The physical domain $r \in [ r_+ , \infty)$ corresponds to the domain $z \in [1, \infty)$; therefore, we impose the Bohr-Sommerfeld quantization condition~\eqref{eq:BS_quant_cond}
to the $B$ cycle.
Similar to the Kerr case~\cite{Aminov:2020yma}, the extremal limit corresponds to $\Lambda_3 \to 0$ and $m_2 \to \infty$ limit for the $N_f = 3$ dictionary, cf.~Eq.~\eqref{eq:DF2SWC_dic1}, where the combination $m_2 \Lambda_3$ remains finite.
This combination can be identified as $\Lambda_2^2 = m_2 \Lambda_3$, when the theory is matched to $N_f = 2$.
We focus on the nonextremal case in this work and leave the study of extremal case for future work.

We emphasize that Eq.~\eqref{eq:DF2SWC_dic1} is not the unique solution because of the
symmetries of the quantum Seiberg-Witten curve~\eqref{eq:Atilde_coefficients}, namely:
\begin{enumerate}
\item the mass exchange $m_1 \leftrightarrow m_2$.
\item the mass-sign flip $m_{1,2} \to - m_{1,2}$.
\item the simultaneous sign flip $m_3 \to -m_3$ and $\Lambda_3 \to - \Lambda_3$.
\end{enumerate}
These three twofold symmetries generate $2^3 = 8$ solutions.
However, only half of the solutions
are distinct since the Nekrasov-Shatashvili free energy $\calF^{\,(3)}$ is symmetric under permutation of the masses $m_i$.
The different choices for the dictionary predict different quasinormal frequencies when the Bohr-Sommerfeld quantization condition~\eqref{eq:BS_quant_cond}
is imposed on $\Pi_B$, and is related to the fact that the parameters $m_{1,2}$ ($m_3$ and $\Lambda_3$) control the boundary conditions imposed at the horizon (at infinity)~\cite{Casals:2021ugr,Ge:2024jdx}.
This also could be an instance of the monodromy ambiguity discussed in Sec.~\ref{sec:sw_ash_conj},
which we leave the detailed study of for the future.

Because the spin-weighted spheroidal harmonic equation~\eqref{eq:S_harmonic_eq}
can also be rewritten in the form of Eq.~\eqref{eq:AGH_SWC}, the eigenvalue problem for $\sw A_{\lm}$ can also be solved using instanton calculations~\cite{Aminov:2020yma}.
The physical domain $u  \in [-1,1]$ is mapped to the domain $z \in [0,1]$ and, therefore, the quantization condition Eq.~\eqref{eq:BS_quant_cond} is imposed on the $A$ cycle.
This is identified as imposing the condition $\mathfrak{a} = \ii (\ell + 1/2 )$, and Ref.~\cite{Aminov:2020yma}
found it leads to the relation,
\begin{align} \label{eq:Alm_inst}
\begin{aligned}
    \sw A_{\lm} &= \ell ( \ell + 1) - s (s + 1) - c^2
    \\
    &\quad + \Lambda_{3} \, {\pd_{\Lambda_{3}}
    \calF^{\,(3)}_{\inst}}(\ii \ell + \ii /2,\, \vecm\,, \Lambda_3 
    ) \vert_{\Lambda_3 = 16 c} \,, \\
    \vecm &= \{ -m,\, -s,\, -s \}\,,
\end{aligned}
\end{align}
which is derived from the Matone relation~\eqref{eq:Matone1};
the eigenvalue $\sw A_{\lm}$ only appears in $E$ in the parameter dictionary Eq.~(4.11) of Ref.~\cite{Aminov:2020yma}.
Unlike the radial case, the three twofold symmetries of the quantum
Seiberg-Witten curve~\eqref{eq:Atilde_coefficients}
lead to the same expression for the angular
eigenvalue $\sw A_{\lm}$.

\subsection{Calculation of quasinormal modes with Leaver's method}
\label{sec:leaver}

To validate the quasinormal frequencies of the Dudley-Finley equation obtained
through Seiberg-Witten theory, we computed the same quantities using
Leaver's method~\cite{Leaver:1985ax}.
Our calculation using Leaver's method is not original and it was first
done in Ref.~\cite{Berti:2005eb}. In this section, we provide a concise
account of this calculation.

The starting point consist in noticing that by imposing quasinormal-mode boundary
conditions on the function $\sw R_{\lm}$, we find the mode functions behave as
\begin{align}
\begin{aligned}
\label{eq:qnm_bcs}
    \lim_{r\,\to\,r_+}    \sw R_{\lm} &\simeq \, (r - r_+)^{-s - \ii \sigma_+},
    \\
    \lim_{r\,\to\,\infty} \sw R_{\lm} &\simeq \, r^{-1-2s+\ii \omega} \, \ee^{\ii \omega r},
\end{aligned}
\end{align}
near the event horizon $r_+$ and at spatial infinity, respectively. Here,
we introduced $\sigma_{+} = [\omega (r_+ - Q^2) - am]/b$, and recall that the
inner and outer horizon locations $r_\pm$ are given by Eq.~\eqref{eq:rpm}.
A solution to Eq.~\eqref{eq:DFeq} satisfying these boundary conditions can be written
in the form~\cite{Leaver:1985ax},
\begin{align}
    \sw R_{\lm} &= \ee^{\ii \omega r}
    \, (r - r_-)^{-1-s+\ii\omega+\ii\sigma_+}
    \, (r - r_+)^{-s-\ii\sigma_+}
    \nonumber \\
                &\quad \times \sum_{n=0}^{\infty} a_{n} \left(\frac{r-r_+}{r-r_-}\right)^{n} \,.
    \label{eq:leaver_df}
\end{align}
Substituting Eq.~\eqref{eq:leaver_df} into Eq.~\eqref{eq:DFeq} yields a three-term recursion relation for the coefficients $a_n$
that we write in a form analogous to Leaver's~\cite{Leaver:1985ax},
\begin{align}
\begin{aligned}
    &\alpha_{0}^{r} \, a_{1} + \beta_{0}^{r} \, a_0 = 0\,, \\
    &\alpha_{n}^{r} \, a_{n+1} + \beta_{n}^{r} \, a_{n} + \gamma_{n}^{r} \, a_{n-1} = 0\,,
    \quad n = 1,\,2,\, \dots
\end{aligned}
\end{align}
The recursion coefficients are
\begingroup
\allowdisplaybreaks
\begin{align} \label{eq:coef_abc_radial}
\begin{aligned}
    \alpha_{n}^{r} &= n^2 + (c_0 + 1) n + c_0 \,,             \\
    \beta_{n}^{r}  &= -2 n^2 + (c_1 + 2) n + c_3\,,           \\
    \gamma_{n}^{r} &= n^2 + (c_2 - 3) n + c_4 - c_2 + 2\,,
\end{aligned}
\end{align}
\endgroup
where we defined the intermediate constants $c_i$,
\begingroup
\allowdisplaybreaks
\begin{subequations}
\label{eq:coef_c}
\begin{align}
    c_0 &= 1 - s - \ii \omega - \frac{2\ii}{b} \left[ \frac{\omega}{2}(1 - 2 Q^2) - a m \right], \\
    c_1 &= -4 + 2 \ii \omega (2 + b) + \frac{4\ii}{b} \left[ \frac{\omega}{2}(1 - 2 Q^2) - a m \right], \\
    c_2 &= s + 3 - 3 \ii \omega - \frac{2\ii}{b} \left[ \frac{\omega}{2}(1 - 2 Q^2) - a m \right], \\
    c_3 &= \omega^2 (4 + 2b - a^2 - 4Q^2) - 2 a m \omega - s - 1 + \ii \omega (2 + b) \nonumber \\
        &\quad - \sw A_{\ell m} + \frac{4 \omega + 2 \ii}{b} \left[ \frac{\omega}{2}(1 - 2 Q^2) - a m \right],\\
    c_4 &= s + 1 - 2 \omega^2 - (2s + 3) \ii \omega - \frac{4 \omega + 2 \ii}{b} \left[ \frac{\omega}{2}(1 - 2 Q^2)
        \right. \nonumber \\
        &\quad\left. \frac{}{}- a m \right] \,.
\end{align}
\end{subequations}
\endgroup
Equations~\eqref{eq:coef_abc_radial}--\eqref{eq:coef_c} are equivalent
to Eq.~(6) in Ref.~\cite{Berti:2005eb}, but we write them here in a way
that makes the Kerr limit ($Q=0$) easier to compare with Eqs.~(24)--(26)
of Leaver~\cite{Leaver:1985ax}.

The series~\eqref{eq:leaver_df} converges and the boundary
conditions~\eqref{eq:qnm_bcs} are satisfied if $\omega$ is a solution to the continued fraction
\begin{equation}
    0 = \beta_0^r
    - \frac{\alpha_0^r \, \gamma_1^r}{\beta_1^r -}
    \, \frac{\alpha_1^r \, \gamma_2^r}{\beta_2^r -}
    \, \frac{\alpha_2^r \, \gamma_3^r}{\beta_3^r -} \cdots,
\end{equation}
for given values of $a$, $Q$, $s$, $m$, and $\sw A_{\ell m}$. The latter is
obtained by solving a similar continued-fraction equation associated to
Eq.~\eqref{eq:S_harmonic_eq}; see Ref.~\cite{Leaver:1985ax}, Eqs.~(20) and (21).

To obtain a quasinormal frequency, the two (radial and angular) continued
fractions must be be satisfied simultaneously. The problem of finding a
quasinormal mode then becomes a root-finding problem.
Our computations were performed in C++. We used Muller's
method~\cite{Muller:MR0083822} to perform root finding, using the
pseudocode from the ``Numerical Recipes,''
Chap.~9.2~\cite{Press2007Numerical}.
We found excellent agreement with the results of Ref.~\cite{Berti:2005eb}.

\section{Numerical results}
\label{sec:nr}

We now compare the values of the quasinormal mode frequencies
calculated from Seiberg-Witten theory, described in Sec.~\ref{sec:canonical_form},
and using the continued fraction method, described in Sec.~\ref{sec:leaver}.
We focus on the fundamental $\ell = 2$ and $m=0$ quasinormal frequencies for illustrative
values of black hole spin $a$ and charge $Q$.

The Seiberg-Witten computations were performed in \texttt{Mathematica}, where the built-in function \texttt{FindRoot} was used to find the roots.
The most time-consuming step is computing the instanton part of the Nekrasov-Shatashvili free energy ${\cal F}_{\text{inst}}^{\,(N_f)}$ from the instanton partition function $Z^{(N_f)}$,
which is obtained as the limit,
\begin{align}
{\cal F}_{\text{inst}}^{\,(N_f)} = - \lim_{\epsilon_2 \to 0} \epsilon_2 \log Z^{(N_f)} \,. \label{eq:NSFEinst}
\end{align}
The bottleneck is due to large cancellations between huge rational expressions, which invalidates the simple substitution $\epsilon_2 = 0$ for taking the limit.

In Table~\ref{tab:comparison}, we show the values of $\omega_{20}$, for three combinations
of $(a, Q)$, namely $(0,0.4)$, $(0.1,0.1)$, $(0.3, 0.1)$, and $(0.48, 0.1)$, computed using both Seiberg-Witten theory (``SW'')
and Leaver's continued fraction method (``CF''). For the former, we report the quasinormal frequencies obtained
by including up to $k=2$, $6$, and $10$ terms in the instanton series in nonperturbative part of the Nekrasov-Shatashvili free energy.
We see that the real and imaginary parts of $\omega_{20}$ in the Seiberg-Witten
calculation oscillate as $k$ increases while approaching the
continued-fraction result. The convergence becomes slower for higher spin
and charge values.
Contrary to Ref.~\cite{Aminov:2020yma}, we have not used a Pad\'e resummation in $\Lambda_3$
to improve the convergence of the instanton contribution to the free energy ${\cal F}_{\rm inst}^{\,(N_f)}$.

\begin{figure}[t]
\includegraphics{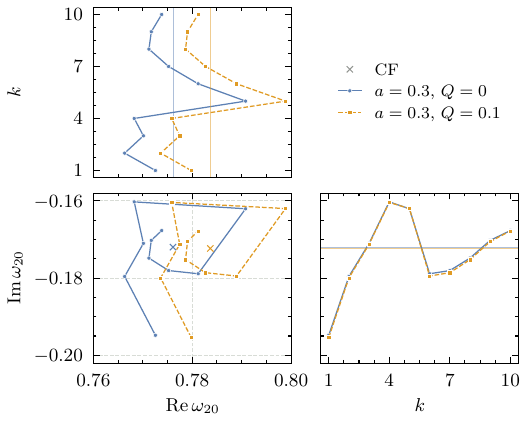}
\caption{Trajectory of the fundamental gravitational quasinormal mode $\omega_{20}$ ($\ell = 2$ and $m = 0$) on the complex plane as we increase the number of terms $k$ in the instanton series from 1 to 10. We show results for the illustrative pairs of spin and charge $(a, Q) = (0.3,0)$ and $(0.3, 0.1)$, represented by the solid and dashed lines, respectively. The crosses indicate the
values of $\omega_{20}$ obtained using the continued fraction method.}
\label{fig:convergence}
\end{figure}

In Fig.~\ref{fig:convergence}, we show the convergence of the Seiberg-Witten
computation towards the result of using Leaver's method, as we increase $k$.
As an illustration, we consider a spin value of $a=0.3$ and charges $Q=0$ (solid line) and $Q=0.1$
(dashed line) for which we obtain the quasinormal frequencies $\omega_{20}=0.776108 - 0.171989\ii$ and $\omega_{20} = 0.783669 - 0.172315\ii$ using Leaver's method, respectively.
The bottom-left panel shows the trajectories of the quasinormal frequencies (parametrized by $k$)
in the complex plane. The result obtained using the continued fraction method is indicated by the
crosses.
The upper-left and bottom-right panels show the real and imaginary parts of $\omega_{20}$
as functions of $k$. In these two panels, the vertical and horizontal lines correspond
to the real and imaginary parts of $\omega_{20}$ obtained using Leaver's method.

\begin{table*}[t]
\begin{tabular}{ c c c c c c c c c c c }
\arrayrulecolor{Gray}
\hline \hline
Method & $\phantom{a}$ & $k$ & $\phantom{a}$ & ($a= 0$, $Q = 0.4$) & $\phantom{a}$ & ($a= 0.1$, $Q = 0.1$) & $\phantom{a}$ & ($a= 0.3$, $Q = 0.1$) & $\phantom{a} $ & ($a= 0.48$, $Q = 0.1$)  \\
\hline
       & & 2              & & $0.875247 - 0.193670 \ii$ & & $0.749438 - 0.184967 \ii$ & & $0.773619 - 0.179956 \ii$ & & $0.829683 - 0.155271 \ii$ \\
SW     & & 6              & & $0.890217 - 0.196240 \ii$ & & $0.759516 - 0.184067 \ii$ & & $0.788964 - 0.179467 \ii$ & & $0.858038 - 0.168951 \ii$ \\
       & & 10              & & $0.884788 - 0.178199 \ii$ & & $0.756254 - 0.175017 \ii$ & & $0.781192 - 0.167787 \ii$ & & $0.835389 - 0.142464 \ii$ \\
\hline
CF     & & $\cdots$       & & $0.886468 - 0.186603 \ii$ & & $0.756664 - 0.178053 \ii$ & & $0.783669 - 0.172315 \ii$ & & $0.850244 - 0.147806 \ii$ \\
\hline \hline
\end{tabular}
\caption{Comparison of fundamental ($n=0$) gravitational quasinormal frequencies $2M\omega_{20}$ ($\ell = 2$ and $m = 0$) of the
Dudley-Finley equation, for illustrative values of spin $a$ and charge $Q$.
The Seiberg-Witten values (SW) converge to continued fraction values (CF) with increasing $k$.
The convergence is slower for the set of parameters $(a,Q)$ closer to extremality.}
\label{tab:comparison}
\end{table*}

\section{Hidden symmetries of the Dudley-Finley equations}
\label{sec:hid_sym}

Following Ref.~\cite{Hatsuda:2020iql} (see Appendix~\ref{app : A}), we can exchange mass parameters to obtain a simpler differential equation that has the same quasinormal mode spectrum as the original Dudley-Finley equation~\eqref{eq:DFeq}.
In particular, the exchange $m_2 \leftrightarrow m_3$ yields the new $A_i'$ coefficients,
\begingroup
\allowdisplaybreaks
\begin{align} \label{eq:DF2SWC_3}
A_0' &= \tfrac{1}{4} - s^2 \,, \nonumber
\\ A_1' &= \sw A_{\ell m} + s(s+2) + a \omega \left( a \omega -  2 m \right) \,, \nonumber
\\ A_2' &= - \sw A_{\ell m} - s(s+1) - a \omega \left( 5 a \omega - 6 m \right) \,,
\\ A_3' &= 4 \omega \left( - a m + 2 a^2 \omega + Q^2 \omega \right) \,, \nonumber
\\ A_4' &= (r_+ - r_-)^2 \omega^2 \,, \nonumber
\end{align}
\endgroup
which are simpler than the original $A_i$ coefficients in Eq.~\eqref{eq:DF2SWC_2}.
We thus have a function $\Psi$ that obeys Eq.~\eqref{eq:AGH_SWC}, but with the coefficients $A'_i$.
Next, we define
\begin{equation}
\Psi(z)=\sqrt{f(z)}\,\phi(z) \,, \quad f(z)=1-z^{-1}\,,
\end{equation}
where $z$ was defined in Eq.~\eqref{eq:r_to_z}.
This results in a simple Regge--Wheeler-like equation for $\phi$, i.e.,
\begin{align}
    \left[f \, \frac{\dd}{\dd z}\left(f\frac{\dd}{\dd z}\right) + \omega^2 - V(z)\right]\phi(z)=0\,,
\label{eq:df_simpler_eq}
\end{align}
with effective potential,
\begin{align}
\begin{aligned}
	V &= f \,
    \left[4(c^2+d^2)+\frac{4 c(m-c)}{z}
    \right. \\
    & \left. \quad + \, \frac{\sw A_{\lm}+s(s+1)-c(2m-c) }{z^2}-\frac{s^2-1}{z^3}\right],
\label{eq:df_simpler_potential}
\end{aligned}
\end{align}
where $c=a\omega$ and $d=Q\omega$. This equation generalizes
the chargeless case presented in Ref.~\cite{Hatsuda:2020iql}, Eqs.~(1) and~(2), to nonzero $Q$.
The potential~\eqref{eq:df_simpler_potential} is real, contrary to
Eq.~\eqref{eq:DFeq}.
In the nonrotating limit $c=0$ and $\sw A_{\lm} = \ell(\ell+1) - s(s-1)$, and Eq.~\eqref{eq:df_simpler_potential} reduces to
\begin{align}
	V &= f \,
    \left[(2 Q \omega)^2+\frac{\ell(\ell+1)}{z^2}-\frac{s^2-1}{z^3}\right].
\label{eq:simple_df_a0}
\end{align}
For $Q \neq 0$, this potential does not recover the gravitoelectromagnetic
perturbation equations of the Reissner-Nordstr\"om solution when we impose either
gravitational or electromagnetic perturbation to be zero; see Ref.~\cite{Berti:2005eb}, Eq.~(20).
This is consistent with the conclusion of Berti and Kokkotas, discussed in Sec.~\ref{sec:DFeq},
that the Dudley-Finley approximation does not completely freeze either perturbations.
Finally, taking the additional limit $Q=0$, we find $z = r$, and Eq.~\eqref{eq:simple_df_a0} reduces to the Regge-Wheeler potential
for generic massless bosonic perturbations of a Schwarzschild black hole.

\section{Conclusions and outlook}
\label{sec:conclusions}

We studied the recently proposed connection between Seiberg-Witten theory and
black hole quasinormal frequencies~\cite{Aminov:2020yma}, in the context of the
gravitoelectromagnetic perturbations of the Kerr-Newman solution in the
Dudley-Finley approximation.
For subextremal black holes and in the parameter space we surveyed, we found
that the lowest damped quasinormal frequencies computed using the
gauge-theoretical tools agree with those obtain using the continued fraction
method~\cite{Leaver:1985ax,Berti:2005eb}.
Our results give further support for the validity of the proposal of Aminov~et~al.~\cite{Aminov:2020yma}. In addition, we obtained a simple
equation that is isospectral to the radial Dudley-Finley equation following Refs.~\cite{Hatsuda:2020iql,Casals:2021ugr}.
Where do we go from here?

One direction would be to study the extremal case, in which the
Dudley-Finley equation maps to the quantum Seiberg-Witten curve for $N_f=2$,
cf.~Eq.\eqref{eq:AGH_SWC_2}. From a physical point of view, however, the problem
is somewhat less interesting as the Dudley-Finley equation becomes a poor
approximation to actual dynamics of gravitoelectromagnetic perturbations in
this limit.

Going beyond the Dudley-Finley approximation, the gravitoelectromagnetic perturbations on Kerr-Newman metric form a system of coupled differential equations, as we described in Sec.~\ref{sec:DFeq}.
If a quantum integrable system that corresponds to a pair of coupled differential equations exists,
and if the system further admits a gauge-theoretic description \`a la Bethe/Gauge correspondence,
the Seiberg-Witten-quasinormal-mode correspondence may provide an alternative method of computing quasinormal modes of Kerr-Newman black holes without resorting to the approximation of Dudley and Finley and complementary to Refs.~\cite{Dias:2015wqa,Dias:2021yju,Dias:2022oqm}.
We leave the exploration of this topic to future work.

Another interesting topic to explore would be to study the asymptotic behavior of quasinormal frequencies through the lens of the Seiberg-Witten-quasinormal-mode correspondence.
It is known that the real part of quasinormal mode frequencies of the Schwarzschild solution asymptotically approach a constant value as the overtone number $n \to \infty$~\cite{Nollert:1993zz,Andersson:1993CQGra,Motl:2002hd,Motl:2003cd,Andersson:2003fh,Natario:2004jd}
which was once conjectured to convey information on the quantum nature of black holes~\cite{York:1983zb,Hod:1998vk};
see~Ref.~\cite{Berti:2009kk}, Sec.~10.1, for a discussion.
Regardless, while the parameter dictionary Eq.~\eqref{eq:DF2SWC_dic1} is unsuitable for studying asymptotic behavior of quasinormal modes  because the expansion parameter $\Lambda_3$ diverges in the asymptotic limit,
it is possible that an alternative expansion exists where the asymptotic limit is convergent.\footnote{We thank Alba Grassi for bringing this point to our attention.}
We leave the study of how such an expansion can be obtained from the
dictionary~\eqref{eq:DF2SWC_dic1} through various dualities satisfied by
supersymmetric gauge theories for future work.

Another direction would be to build on the extension explored by Ref.~\cite{Bonelli:2021uvf},
i.e., applying the Alday-Gaiotto-Tachikawa correspondence~\cite{Alday:2009aq} to construct the solutions to the perturbation equations
which can be used, e.g., to obtain resummed expressions for several observables in perturbation theory.
For Kerr black holes, this has been used to obtain post-Minkowskian-resummed expressions for the graybody factor~\cite{Bonelli:2021uvf}, or the gravitational Compton amplitude~\cite{Bautista:2023sdf,Bautista:2024emt}.
The extension of Seiberg-Witten correspondence to the Kerr-Newman case
can now be used to extend these studies for the case of charged-spinning black holes
in the Dudley-Finley approximation.
Subsequently, the resummed Compton amplitude may also be used to obtain resummed expressions for the tidal response following earlier works~\cite{Saketh:2022xjb,Saketh:2023bul,Ivanov:2022qqt,Saketh:2024juq,Chakrabarti:2013lua,Creci:2023cfx,Bautista:2022wjf,Bautista:2024emt,Bautista:2024agp}.
In particular, this could help test the near-far factorization in the Compton amplitude~\cite{Ivanov:2022qqt,Ivanov:2024sds}, look for unique features in their tidal response, and fix coefficients in an effective worldline action for charged-spinning black holes. Such works could further highlight the usefulness of the Seiberg-Witten correspondence towards analytical studies in classical gravitational physics.

Finally, there are some improvements that can be implemented on the Seiberg-Witten side.
The computational bottleneck on the quantum Seiberg-Witten curve approach is evaluation of the Nekrasov-Shatashvili free energy.
While the formulas for the instanton partition function $Z^{(N_f)}$ can be evaluated very efficiently,
the limiting procedure required to compute the Nekrasov-Shatashvili free energy~\eqref{eq:NSFEinst} is very slow in general.
As we remarked in Sec.~\ref{sec:nr}, this is caused by the large number of cancellations between coefficients of $Z^{(N_f)}$, which are rational functions.
We can expect to speed up the computation of the Nekrasov-Shatashvili free energy by using specialized techniques for simplifying rational functions,
such as functional reconstruction using finite fields. This technique has been proven powerful in Feynman diagram calculations~\cite{vonManteuffel:2014ixa,Peraro:2016wsq}.
Another possibility is to use continued-fraction methods for computing the Nekrasov-Shatashvili free energy~\cite{Poghosyan:2020zzg} as advocated recently in Ref.~\cite{Cipriani:2025ikx}.
Such an improvement would be necessary to study quantum Seiberg-Witten curves at higher orders in the instanton series.

\section*{Acknowledgments}

The authors would like to thank Gleb Aminov, Alba Grassi, and Yasuyuki Hatsuda
for valuable discussions and sharing the codes used for the calculations presented in
Ref.~\cite{Aminov:2020yma} which we used to validate some of our results.
We thank the members of the interdepartmental ``\texttt{bh-hep-th}'' reading
group at the Max Planck Institute for Gravitational Physics, in particular
Nikita Misuna, Raj Patil, and Giovanni Tambalo, for the many lively discussions.
We also acknowledge discussions with Rita Teixeira da Costa.
J.-W.K would like to thank Saebyeok Jeong and Ki-Hong Lee for discussions and
sharing codes for checking some of our calculations.
H.O.S. acknowledges funding from the Deutsche Forschungsgemeinschaft
(DFG)~-~Project No.:~386119226.

\appendix
\section{Point spectrum and mass exchange}
\label{app : A}

In Ref.~\cite{Casals:2021ugr}, the symmetry of the point spectrum for Eq.~\eqref{eq:AGH_SWC} with respect to swapping of $m_1$, $m_2$, $m_3$ for appropriate boundary conditions was shown. This is expected from the invariance of the Seiberg-Witten theory under the exchange, but not obvious from the differential equation itself. The symmetry of the point spectrum under exchange of $m_1$ and $m_2$ is obvious from the equation itself, as the coefficients $A_{i}$ are all invariant for $m_1\leftrightarrow m_2$. This is not however clear for the exchange of $m_{1,2}$ and $m_3$. In this appendix, we thus briefly repeat the presentation used in Ref.~\cite{Casals:2021ugr} justifying the symmetry of the point spectrum under exchange of $m_1$, $m_2$, and $m_3$. This is mainly to point out the sign-choices required to ensure the right boundary conditions, and its relation to the dictionary Eq.~\eqref{eq:DF2SWC_dic1} derived by us.

We first identify and remove the asymptotic behavior in the limits $z\rightarrow 1$ and $z\rightarrow \infty$, corresponding to the quasinormal mode boundary conditions when the Dudley-Finley dictionary Eq.~\eqref{eq:DF2SWC_dic1} is used. For $z\rightarrow 1$,
we find that the solution of Eq.~\eqref{eq:AGH_SWC} is
\begin{equation}
	\Psi(z) \sim (z-1)^{\frac{1}{2}[1 \pm (m_1+m_2)]}\,.
    \label{eq:psi_z}
\end{equation}
By choosing the plus sign in the equation above, and substituting the dictionary~\eqref{eq:DF2SWC_dic1} [along with Eq.~\eqref{eq:DF2SWC_1}] in Eq.~\eqref{eq:psi_z}, we obtain
\begin{alignat}{3}
\label{eq:app_radial}
	R(r) \sim (r-r_+)^{-s-\ii \sigma_+}, \quad \Psi(z) \sim (z-1)^{\frac{1-s}{2}- \ii \sigma_+ },
\end{alignat}
where $\sigma_+= [\omega (r_+-Q^2)-a m]/b$.
This corresponds to the appropriate incoming boundary conditions at the horizon. Similarly, in the limit $z\rightarrow \infty$, Eq.~\eqref{eq:AGH_SWC} becomes
\begin{alignat}{3}
	\left[\frac{\dd^2}{\dd z^2}-\left(\frac{\Lambda_3^2}{64}+\frac{m_3\Lambda_3}{4z}\right)\right]\Psi(z)=0\,,
    \quad z \to \infty
    \label{eq:dpsi_dz}
\end{alignat}
with solutions $\Psi(z)\sim \ee^{\,\pm \Lambda_3 z/ 8} \, z^{\,\pm m_3}$. Choosing the minus sign here along with the dictionary in  Eq.~\eqref{eq:DF2SWC_dic1} yields the solution $R(r)\sim  \ee^{\ii \omega r}$, as $r\rightarrow \infty$, which corresponds to outgoing boundary conditions at spatial infinity.
We now substitute
\begin{align}
\begin{aligned}
    \Psi(z) &=
    \exp\left[-\tfrac{1}{8}\Lambda_3 (z-1)\right]
    \, z^{\frac{1}{2}(-2 m_3-1-m_1-m_2)}
    \\
    &\quad (z-1)^{\frac{1}{2}(1+m_1+m_2)}
    \, g(z) \,,
\end{aligned}
\end{align}
in Eq.~\eqref{eq:dpsi_dz}, such that the function $g$ is regular at
$z\rightarrow 1$ and $z\rightarrow\infty$ for quasinormal mode frequencies.
We obtain the following equation for $g$:
\begin{align} \label{eq:tex}
    &\left\{
    z(z-1) \, \frac{\dd^2}{\dd z^2}+[B_1z(z-1)+B_2(z-1)+B_3] \, \frac{\dd}{\dd z} \right.
    \nonumber
    \\
      &\left. +B_4 \, \left(1 - \frac{1}{z} \right) +B_5 \right\} \, g(z)=0 \,,
\end{align}
where the coefficients $B_i$ ($i = 1, \dots, 5$) are related to those given in Ref.~\cite{Casals:2021ugr} by the
transformations $\Lambda_3\rightarrow -\Lambda_3$ and $m_i\rightarrow -m_i$.
Reference~\cite{Casals:2021ugr} went on to show that Eq.~\eqref{eq:tex} has a point spectrum that is invariant under $m_i\leftrightarrow m_j$. Since the transformation relating our $B_i$ to theirs commutes with the operation of  swapping the masses, the rest of the argument for the invariance of the point spectrum goes through unchanged in a similar manner to Ref.~\cite{Casals:2021ugr}.

\bibliography{biblio}

\end{document}